\documentclass[aip,pop,amsfonts,amsmath,amssymb,reprint]{revtex4-2}
\pdfoutput=1 

\usepackage{graphicx}

\usepackage[english]{babel}
\usepackage[utf8]{inputenc}
\usepackage[pdftitle={Article}, pdfauthor={Author}]{hyperref} 

\begin{document}


\title{Enhancement of axial magnetic field generation during relativistic self-channeling of laser radiation propagating along thin films} 

\author{Vitalia A. Kuleshova}
\affiliation{Lomonosov Moscow State University, Moscow, Russia}
\author{Artem V. Korzhimanov}
\email[Correspondence email address: ]{artem.korzhimanov@ipfran.ru}
\affiliation{Gaponov--Grekhov Institute of Applied Physics of the Russian Academy of Sciences, Nizhny Novgorod, Russia}
\affiliation{Lobachevsky State University of Nizhny Novgorod, Nizhny Novgorod, Russia}
\date{\today}

\begin{abstract}
Within the framework of a stationary model of relativistic self-focusing, it is shown that the addition of a thin layer of denser plasma on the propagation axis of a circularly polarized beam increases the axial magnetic field generated by it via the inverse Faraday effect by more than a factor of 10 compared to the case of a homogeneous plasma. The magnitude of the axial field in this case can exceed the magnetic field of the laser wave, which makes it possible, at radiation intensities $\sim 10^{26}$~W/cm$^2$ potentially achievable in the near future, to approach the level of teragauss quasi-stationary fields.
\end{abstract}

\maketitle 

\section{\label{sec:intro}Introduction}

Modern sources of ultrashort ultrahigh-power laser pulses are among the most efficient means of concentrating energy, making it possible to reach radiation intensities of up to $10^{23}$~W/cm$^2$~\cite{korzhimanov_Ufn_2011,danson_HPLSE_2015,yoon_O_2021}. Multi-channel systems with a total power of 100 petawatts and above, currently being designed~\cite{khazanov_HPLSE_2023}, will potentially allow, with optimal focusing, to exceed an intensity of $10^{26}$~W/cm$^2$~\cite{bulanov_AO_2025,sidnev_JOSAB_2025}. The interaction of such laser pulses with matter leads to efficient acceleration of particles to high energies~\cite{esarey_RMP_2009,daido_RPP_2012,macchi_RMP_2013}, generation of bright flashes of X-ray and gamma radiation~\cite{corde_RMP_2013,albert_PPCF_2016}, and can be used to produce dense beams of neutrons~\cite{yogo_EPJA_2023}, positrons~\cite{gonoskov_RMP_2022,kostyukov_Ke_2024}, muons~\cite{zhang_NP_2025}, as well as to generate powerful pulses in the low-frequency~\cite{consoli_HPLSE_2020}, terahertz~\cite{gopal_PRL_2013}, and mid-infrared~\cite{nie_NP_2018,mikheytsev_MRE_2023,lei_PRL_2025} ranges.

One of the research directions in the field of applications of ultrahigh-power laser pulses is the generation of quasi-stationary magnetic fields capable of existing on timescales exceeding the pulse duration. Experimentally, magnetic fields of the order of several megagauss have been detected, arising when laser pulses with energies of hundreds of joules and above irradiate targets with broken rotational symmetry (microcoils, microsolennoids, etc.)~\cite{law_APL_2016,santos_PP_2018,gao_APL_2025}. Such fields exist on nanosecond timescales in volumes of the order of a cubic millimeter. Magnetic fields of the same magnitude have also been measured in transparent plasma during the propagation of a circularly polarized picosecond pulse through it~\cite{najmudin_PRL_2001}. On the surfaces of solid targets irradiated by femtosecond pulses, fields ranging from several megagauss~\cite{choudhary_FPP_2025} to tens of megagauss~\cite{schumaker_PRL_2013} have been measured. When thin targets were irradiated from the rear side, magnetic fields of up to hundreds of megagauss were indirectly detected~\cite{nakatsutsumi_NC_2018}. Such fields typically exist on picosecond timescales in volumes of the order of hundreds of cubic microns.

At the same time, the amplitude of the magnetic field in the laser wave is $B = 91.6(I_{18})^\frac12$~MG, where $I_{18}$ is the radiation intensity expressed in units of $10^{18}$~W/cm$^2$, and thus, at intensities typical for experiments of the order of $10^{18}$--$10^{20}$~W/cm$^2$, the amplitude of the magnetic field in the wave amounts to hundreds of megagauss, which significantly exceeds the measured values of quasi-stationary fields. In this regard, the question arises whether it is possible to arrange the interaction such that the oscillating field of the wave and the generated quasi-stationary field are of the same order of magnitude.

Several promising approaches to the generation of magnetic fields in laser-plasma interactions are discussed in the literature, among which one can distinguish the generation of azimuthal fields by longitudinal currents of electrons accelerated by the laser pulse~\cite{pukhov_PRL_1996,nakatsutsumi_NC_2018} and the inverse Faraday effect during the propagation of circularly polarized radiation~\cite{kim_PRL_2002,naseri_PP_2010a,liseykina_NJP_2016} or radiation carrying orbital angular momentum~\cite{lecz_SR_2016,nuter_PRE_2018,shi_PRL_2023} in plasma. The generated fields can also be enhanced by target implosion~\cite{murakami_SR_2020}, through pinching of the excited currents~\cite{kaymak_PRL_2016}, and by using cluster and nanostructured targets~\cite{andreev_Zh_2025}.

The main feature here is that at radiation intensities above $10^{18}$~W/cm$^2$ with a wavelength of about 1~$\mu$m, the kinetic energy of electrons becomes comparable to their rest energy and their velocity approaches the speed of light. This leads to the fact that the magnitude of the generated fields in optimal regimes is determined almost exclusively by the power of the laser radiation. Indeed, by order of magnitude, the field generated by relativistic currents can be estimated as $B\sim jL\sim eN_ecL$, where $j$ is the characteristic density of the induced current in the target, $L$ is the characteristic size of the localization of this current, $e$ is the elementary charge, $N_e$ is the electron concentration, and $c$ is the speed of light. At the same time, efficient generation of currents is possible only in transparent plasma, since in opaque plasma the axial current is excited inefficiently, while the azimuthal current is localized in a thin skin layer. Taking into account relativistic nonlinearity, this means that the most efficient is the use of plasma with near-critical concentration: $N_e\sim N_{cr}\sim \sqrt{1+a_0^2}\pi m_ec^2/(e^2\lambda^2) \approx 1.1a_0\lambda_{\mu m}^{-2}\times 10^{21}$~cm$^{-3}$, where $a_0 = e\lambda/m_ec^2(2I/\pi c)^\frac12 = \lambda_{\mu m}\sqrt{I_{18}/1.38}$ is the dimensionless amplitude of the laser pulse field, $m_e$ is the electron mass, $\lambda$ is the radiation wavelength, and $\lambda_{\mu m}$ is the radiation wavelength measured in microns. In the case of a homogeneous plasma, the diameter of the laser beam $w$ serves as the size $L$, and thus the generated magnetic field has the scaling $B\propto a_0 w\propto \sqrt{P}$, where $P = Iw^2$ is the power of the laser beam. Thus, reducing the beam diameter in order to increase the intensity does not lead to an increase in the generated magnetic field, since the localization region of the induced currents decreases proportionally.

In this work, we propose a relatively universal way to circumvent this limitation, which consists of adding to the interaction region a thin layer of overdense plasma --- of the order of the skin depth in thickness --- oriented along the direction of propagation of the laser pulse. Such a layer, on the one hand, does not lead to a significant change in the optical properties of the medium and to reflection of the laser radiation, and on the other hand, it allows one to increase the number of electrons in the interaction region and thereby enhance the induced currents and the generated quasi-stationary fields. Such a layer can be a film or foil of submicron thickness, or a wire or rod of submicron diameter.

To demonstrate the effect, we consider a specific method of laser-plasma generation of a quasi-stationary magnetic field: the inverse Faraday effect during self-channeling of relativistically strong circularly polarized laser radiation~\cite{kim_PRL_2002}. Despite the limited nature of the model used, its conclusions regarding the influence of submicron dense plasma layers on the generation of magnetic fields will be fairly universal in character. Our aim is to show that the application of this idea makes it possible to obtain quasi-stationary magnetic fields with a magnitude comparable to the oscillating magnetic field of the laser wave, which, at intensities of the order of $10^{26}$~W/cm$^2$, would allow one to reach fields of the order of 1 TGa. At such fields, the magnetic component of the Lorentz force acting on electrons in atoms becomes comparable to the intraatomic electric field, leading to a significant distortion of the electronic states and the atomic emission spectrum~\cite{Liberman_UFN_1995,lai_RMP_2001}. The study of atoms in such fields is of interest for astrophysics and, in particular, for the physics of compact stellar objects such as neutron stars~\cite{beskin_SSR_2015}.

The generation of magnetized plasma is also of interest from the standpoint of studying astrophysically relevant regimes of interaction of magnetized plasma flows~\cite{bulanov_PPR_2015}. In particular, ultrahigh-power laser pulses are capable of generating strongly magnetized cold relativistic plasma that is inaccessible by other methods~\cite{korzhimanov_AS_2021,sladkov_Ke_2023}.

\section{Model of axial magnetic field generation via the inverse Faraday effect during relativistic self-channeling}

Let us consider the propagation of a circularly polarized relativistically intense laser pulse in plasma. As is known, in this case the effect of self-focusing can be observed, which is caused by the dependence of the plasma dielectric permittivity on the local wave amplitude due to relativistic and strictional (ponderomotive) nonlinearities~\cite{talanov_IvR_1964,litvak_Zh_1969a,max_PRL_1974,bychenkov_Ke_2024,bychenkov_PvZh_2024,kovalev_Zh_2025}. When the power of the laser beam exceeds the critical power for self-focusing $P_c = 2m_e^2c^5/e^2\cdot N_{cr}^0/N_e \approx 17.4 N_{cr}^0/N_e$~GW, where $N_{cr}^0= \pi m_ec^2/e^2\lambda^2 = 1.1\lambda_{\mu m}^{-2}\times 10^{21}$~cm$^{-3}$ is the critical concentration in the linear regime, propagation of the laser beam in a quasi-stationary regime becomes possible, in which self-focusing exactly compensates for the diffraction divergence and, neglecting losses, the beam diameter remains unchanged during propagation. Such a regime is commonly referred to as self-channeling.

In one of the simplest models, the transverse structure of the laser beam and the self-consistent plasma channel can be obtained neglecting ion motion and harmonic generation within the framework of the cold hydrodynamics equations and Maxwell's equations~\cite{sun_PF_1987,chen_PFBPP_1993}. A well-known feature of such solutions, however, is their discontinuous nature, associated with the neglect of electron pressure, while the correct procedure for constructing the solution requires imposing an additional condition of global quasineutrality, which complicates the solution of the differential equations~\cite{cattani_PRE_2001,kim_PRE_2002}. Alternatively, the problem of discontinuous solutions can be circumvented by regularizing the equations by adding to the hydrodynamic equations a term proportional to the density gradient~\cite{feit_PRE_1998}. This increases the order of the system of differential equations to be solved, but ensures the smoothness of the solutions~\cite{korzhimanov_EPJD_2009}.

Note that, as will be seen below, we will be interested, among other things, in solutions with two humps of the laser wave in the transverse direction. As was shown in Refs.~\cite{cattani_PRE_2001,kim_PRE_2002}, in the cold hydrodynamics approximation, in this case two discontinuities in the electron density also arise, as well as a free parameter --- the width of the cavitation region between the discontinuities --- whose value can be arbitrary within a certain range. The presence of such a parameter, together with the need to match the solutions of the nonlinear equations in the two spatial regions by boundary conditions, complicates the application of the cold hydrodynamics model; therefore, in this work we will use the approach with regularization of the equations, which is free from such difficulties.

In planar two-dimensional geometry, a stationary laser-plasma channel is described by the following system of dimensionless equations~\cite{feit_PRE_1998,cattani_PRE_2001,korzhimanov_EPJD_2009}:
\begin{eqnarray}
    \frac{d^2a}{d\xi^2} &+& a = \alpha\frac{na}{\sqrt{1+a^2}} \label{eq:helmholtz} \\
    \frac{d f}{d\xi} &=& \alpha\left(n_i(\xi) - n\right) \label{eq:gauss} \\
    \frac{\mu}{n}\frac{dn}{d\xi} &+& f + \frac{d\sqrt{1+a^2}}{d\xi} = 0, \label{eq:balance}
\end{eqnarray}
where the vector potential of the electromagnetic wave is assumed to have the form $\mathbf{A} = m_ec^2/e\cdot a(\xi)(\mathbf{x_0} + i\mathbf{y_0})\exp \{i(hz-\omega t)\}$, $\xi = \sqrt{k^2-h^2} x$, $k = \omega/c$, $\mathbf{x_0}$, $\mathbf{y_0}$ are the unit vectors of the corresponding axes, $h$ is the propagation constant, $\omega$ is the wave frequency,
$$
\alpha = \frac{4\pi e^2 N_0}{m_e\left(\omega^2 - h^2c^2\right)} = \frac{N_0/N_{cr}^0}{1-h^2/k^2},
$$
$N_0$ is some characteristic particle concentration in the plasma, $n=N_e/N_0$, $n_i=N_i/N_0$ are the electron and ion concentrations, respectively, with the function $n_i(\xi)$ assumed to be given and characterizing the background plasma density profile, $f = eE_x/(m_ec^2\sqrt{k^2-h^2})$ is the electrostatic charge-separation field in the plasma, and $\mu$ is the regularization parameter, which in the nonrelativistic limit has the meaning of the electron temperature normalized to the rest energy. In the relativistic case, the parameter $\mu$ does not have a specific physical meaning, and its introduction should be regarded solely as a mathematical technique equivalent to introducing a small parameter for the highest derivative.

Equation (\ref{eq:helmholtz}) is a consequence of the Helmholtz equation for the transverse modes of the waveguide, equation (\ref{eq:gauss}) represents Gauss's law for the electrostatic field, and equation (\ref{eq:balance}) expresses the balance of forces acting on the electrons: the pressure force, the electrostatic force, and the ponderomotive force. It is convenient to rewrite the system (\ref{eq:helmholtz})--(\ref{eq:balance}) as two second-order equations for the potentials. To this end, we introduce the electrostatic potential $\phi$: $f = -d\phi/d\xi$ and integrate equation (\ref{eq:balance}):
\begin{equation}
    n = \exp\frac{\sqrt{1+a^2} - \phi + 1}{\mu},
\end{equation}
where the integration constant is chosen such that in the region where the potentials vanish $a=\phi=0$, the density is $n=1$. From this expression it is explicitly seen that for $\mu >0$ the quantity $n$ is everywhere nonnegative. Taking into account equation (\ref{eq:gauss}) and zero boundary conditions, this ensures that the total charge of the plasma is zero, in contrast to the case $\mu=0$.
As a result, the two remaining equations take the form:
\begin{eqnarray}
    \frac{d^2a}{d\xi^2} &=& a\left(\frac{\alpha}{\sqrt{1+a^2}}\exp\frac{\sqrt{1+a^2} - \phi + 1}{\mu} - 1\right) \label{eq:a} \\
    \frac{d^2\phi}{d\xi^2} &=& \alpha\left(\exp\frac{\sqrt{1+a^2} - \phi + 1}{\mu} - n_i(\xi)\right) \label{eq:phi}
\end{eqnarray}
Note that this system was previously analyzed for the case of longitudinal modes in Ref.~\cite{korzhimanov_EPJD_2009}.

For circularly polarized radiation in an inhomogeneous plasma, the effect of generation of a longitudinal (axial) quasi-stationary magnetic field is observed due to the inverse Faraday effect~\cite{najmudin_PRL_2001}. In the model of relativistic self-channeling, the magnitude of this field can be determined from the following equation for the zeroth harmonic of the azimuthal component of the vector potential $A_\varphi$~\cite{kim_PRL_2002}:
\begin{equation}
    \frac{d^2a_\varphi}{d\xi^2} - \frac{\alpha n}{\sqrt{1+a^2}}a_\varphi = \sqrt{\frac{\alpha N_0}{N_{cr}}}\frac{a^2}{2\sqrt{1+a^2}}\frac{d}{d\xi}\frac{n}{\sqrt{1+a^2}}, \label{eq:a_phi}
\end{equation}
where $a_\varphi = eA_\varphi/m_ec^2$. It is assumed that $a_\varphi \ll a$, and this equation is solved independently of the system (\ref{eq:a})--(\ref{eq:phi}) with the functions $n(\xi)$ and $a(\xi)$ obtained from it. The magnitude of the axial magnetic field can be calculated as:
\begin{equation}
    B = \frac{m_ec\omega}{e}\sqrt{\frac{N_0}{\alpha N_{cr}}}\frac{da_\varphi}{d\xi},
\end{equation}
where $B_{rel} = m_ec\omega/e = 107/\lambda_{\mu m}$~MG.

Note that although, as is well known~\cite{abdullaev_PvZh_1981,gorbunov_Zh_1998,frolov_Fp_2004} and as directly follows from equation (\ref{eq:a_phi}), the source of the magnetic field in the inverse Faraday effect in plasma is currents determined by the gradient of the quantity $n/\sqrt{1+a^2}$, i.e. gradients of the laser field and density, rather than by their magnitudes, in the case of interest to us these currents are localized in a layer of thickness smaller than the skin depth. In this case, in equation (\ref{eq:a_phi}) one can neglect the second term, as well as the inhomogeneity of the quantity $a(\xi)$, and then the field will be determined only by the density jump and will not depend on the details of its distribution and the local value of the gradient. The influence of the plasma density profile in this case will be exclusively indirect --- through the change in the optical properties of the plasma and, consequently, in the quantity $a(\xi)$ at the location of the density jump.

The system of equations (\ref{eq:a})--(\ref{eq:a_phi}) is the subject of study in this work. We will consider its stationary solutions in infinite space with the condition that all potentials decay to zero at infinity. In this case, the solutions of the system are determined by two parameters, $\alpha$ and $N_0/N_{cr}$. A known difficulty is that in the physical formulation, the externally specified parameter is not $\alpha$, but the power of the laser pulse, which, for a given plasma density, determines the shape of the plasma channel and the corresponding propagation constant $h$, which in turn determines the value of $\alpha$. In this regard, to construct the dependences at fixed radiation power, we first constructed dependences on the parameter $\alpha$ and for each obtained solution determined the linear power of the laser beam:
\begin{equation}
    Q = \frac{m_e^2c^4\omega}{e^2}\sqrt{\frac{\alpha N_{cr}}{N_0}}\int\limits_{-\infty}^{\infty}a^2d\xi, \label{eq:power}
\end{equation}
and then performed an interpolation procedure for the desired linear power. For numerical estimates, we note that $Q_{rel} = m_e^2c^4\omega/e^2 = 54.7/\lambda_{\mu m}$~GW/$\mu$m.

Note also that the system (\ref{eq:a})--(\ref{eq:a_phi}), generally speaking, can have a discrete spectrum of solutions (modes)~\cite{cattani_PRE_2001,korzhimanov_EPJD_2009}, but within the framework of this study we restricted ourselves to the analysis of only the fundamental mode --- with the lowest linear power $Q$. The search for such a mode was carried out using the standard collocation method with the \textit{SciPy} library~\cite{virtanen_NM_2020}.

\begin{figure*}[t]
   \centering
   \includegraphics[width=\linewidth]{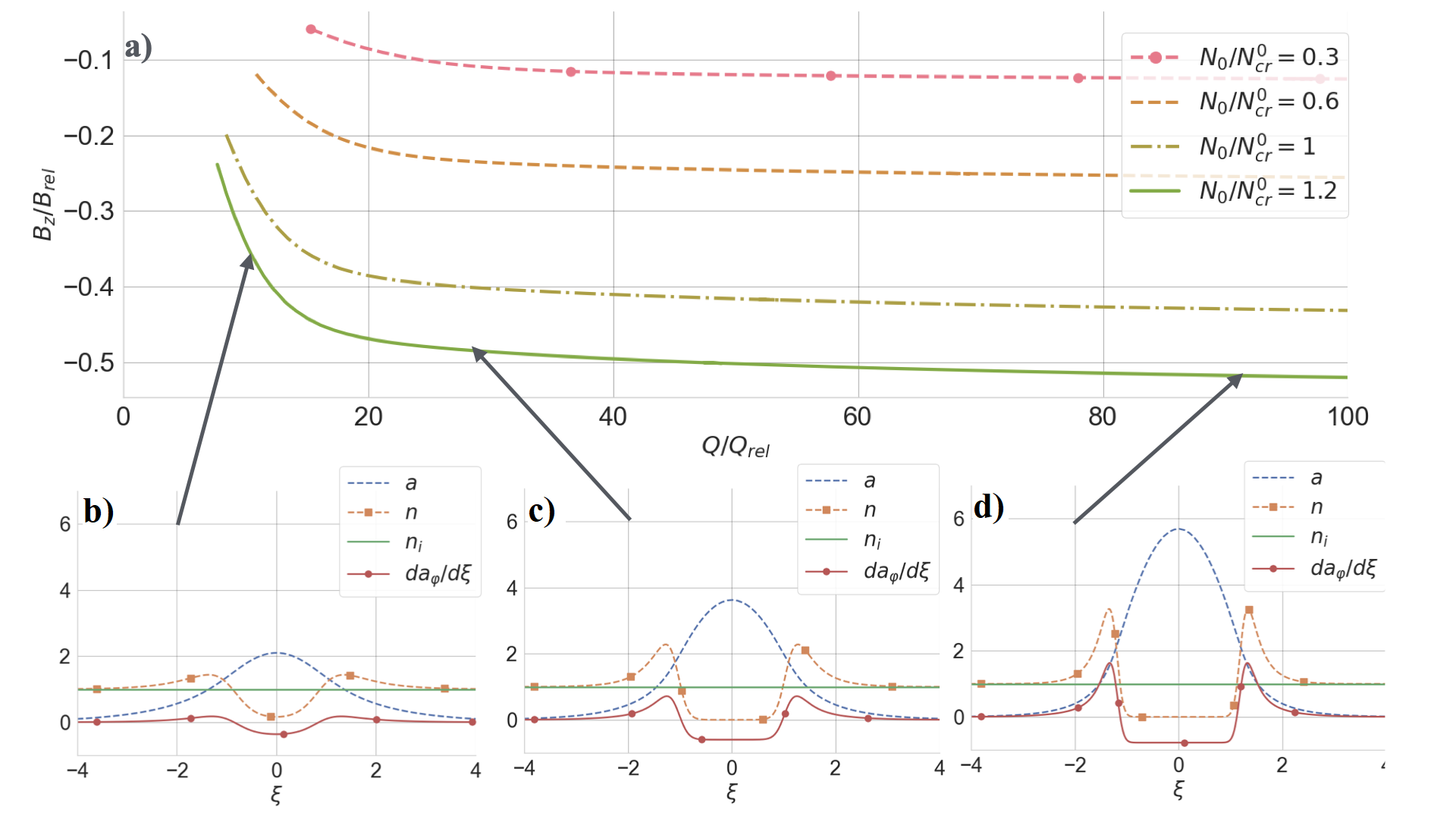}
   \caption{(a) Dependence of the axial magnetic field magnitude on the axis of the laser beam on its power during self-channeling in an initially homogeneous plasma at a fixed plasma density. (b)--(d) Distributions of the laser field, electron density, ion density, and axial magnetic field at $Q/Q_{rel}=10$, $30$, $90$, respectively, and $N_0 = 1.2N_{cr}^0$ (marked by arrows in the inset (a)). The plots are constructed for $\mu=0.03$.
   }
   \label{fig:homo_vsQ}
\end{figure*}

\section{Analysis of the dependence of the axial magnetic field on the system parameters}

First, the dependence of the generated axial magnetic field on the laser radiation power at a fixed plasma density was investigated for the case of a homogeneous ion density distribution $n_i(\xi)\equiv 1$. The result is shown in Fig.~\ref{fig:homo_vsQ}. Note that for the chosen laser pulse polarization, the axial magnetic field on the beam axis has a negative projection onto the $z$ axis.

From the analysis of Fig.~\ref{fig:homo_vsQ} (a), it can be seen that solutions exist only at powers above some critical value, and for a small excess above this power the generated magnetic field grows relatively rapidly; however, then this growth noticeably slows down.

In Figs.~\ref{fig:homo_vsQ} (b)-(d), the typical distributions of the laser field, electron density, and axial magnetic field are shown for three cases: in the regime of rapid field growth, in the transition region, and in the regime of slow growth. From their analysis, it can be seen that the field grows rapidly in the case when complete cavitation (expulsion of electrons) of the plasma channel does not occur. In this region, as the power increases, the amplitude of the laser field on the axis grows with increasing power, which leads to an increase in the magnetic field. However, at a certain power, complete expulsion of electrons from the plasma channel occurs; in this case, the plasma can maintain self-focusing only by increasing the channel width, but not by increasing the field amplitude within it. In this parameter region, the magnetic field on the axis is determined almost exclusively by the structure of the channel boundaries, which depends weakly on the power, so the growth of the magnetic field with increasing radiation power stops. Such behavior for the magnetic field on the beam axis was previously observed in Ref.~\cite{kim_PRL_2002}, in which a comparable value of the limiting field $\sim 0.3B_{rel}N_0/N_{cr}^0$ was obtained ($\sim 0.4B_{rel}N_0/N_{cr}^0$ in our case). The quantitative difference is mainly explained by the fact that in our work a two-dimensional approximation was used, while in Ref.~\cite{kim_PRL_2002} a three-dimensional (azimuthally symmetric) one was used. Somewhat larger, but of the same order, fields ($\sim 0.02 B_{rel}$ at $N_0=0.02N_{cr}^0$ in the full cavitation regime) were obtained in kinetic numerical simulations in Ref.~\cite{naseri_PP_2010a}. The difference is possibly related to the fact that in the simulations the steady-state self-focusing regime was not reached due to the too short propagation path of the radiation. Nevertheless, in that work, a weak dependence of the generated magnetic field magnitude on the radiation power was also observed.

\begin{figure}
   \centering
   \includegraphics[width=\linewidth]{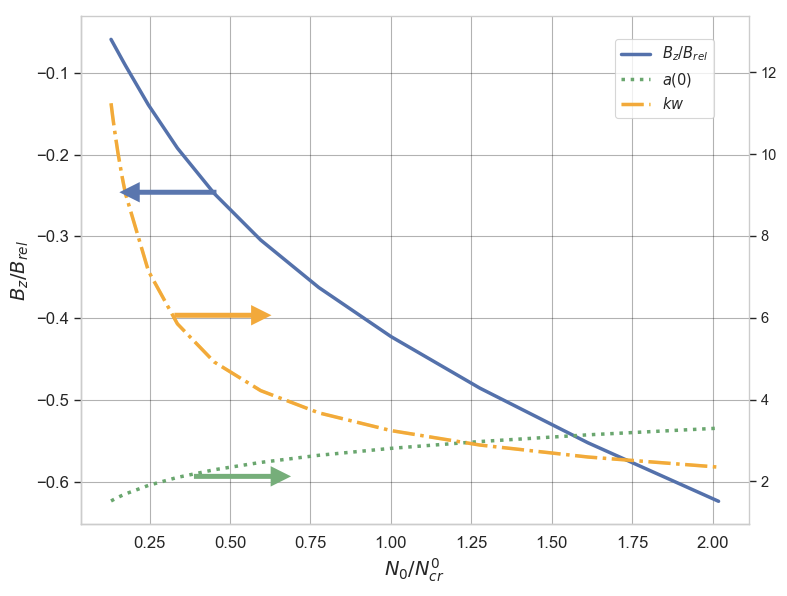}
   \caption{Dependences of the axial magnetic field magnitude on the axis of the laser beam $B_z$, the laser field amplitude at the same point $a(0)$, and the beam width $w$ at half-maximum during self-channeling in an initially homogeneous plasma on the plasma density at a fixed linear power $Q=20 Q_{rel}$. The dependences are plotted for $\mu=0.03$.
   }
   \label{fig:homo_vsN}
\end{figure}

The transition to the regime of incomplete cavitation at a given radiation power can be achieved by increasing the plasma density, which should lead to an enhancement of the generated magnetic field. The corresponding dependence is shown in Fig.~\ref{fig:homo_vsN}.

As can be seen, with increasing plasma density, a growth of the magnetic field is indeed observed. Also shown is how the laser field amplitude and the beam width change in this case. It is seen that in a relatively dense plasma ($N_0 > 0.5 N_{cr}^0$ at the given parameters), the pulse amplitude and the beam width almost cease to change with increasing density. Thus, the growth of the field is provided mainly by the increase in density, which increases the number of electrons involved in motion by the laser field and, accordingly, the currents generated in the plasma.

However, there is a significant limitation: at a certain density, the curve breaks off, since at high concentrations it turns out that $N_0 > \alpha N_{cr}^0$, and the propagation constant $h$ becomes imaginary, which corresponds to a wave decaying along the $z$ axis --- the plasma becomes opaque (overdense). Note that even in this limit, the generated magnetic field is several times smaller in magnitude than the amplitude of the laser radiation field.

Thus, the solutions are such that to increase the amplitude of the laser field, one needs to reduce the width of the plasma channel, which requires an increase in the plasma density; however, this leads to cavitation of the channel and expulsion of electrons from the region of the strongest laser field, which limits the currents excited in the plasma and the generated magnetic field. To overcome this contradiction, we propose to use a plasma with a transversely profiled density of the form:
\begin{equation}
    n_i(\xi) = 1 + (n_{peak} - 1)e^{-\frac{\xi^2}{b^2}}. \label{eq:profile}
\end{equation}
The quantity $n_{peak} > 1$ characterizes the plasma density on the beam axis, and $b$ is the width of the density spike, with the assumption that $b < 1/\sqrt{n_{peak}}$, so that it is thinner than the skin depth and does not introduce significant perturbations into the transverse structure of the laser beam. In this case, the background plasma density controls the width of the laser beam, while the product $n_{peak}b$ controls the number of electrons in the region of the strong field.

\begin{figure*}
   \centering
   \includegraphics[width=\linewidth]{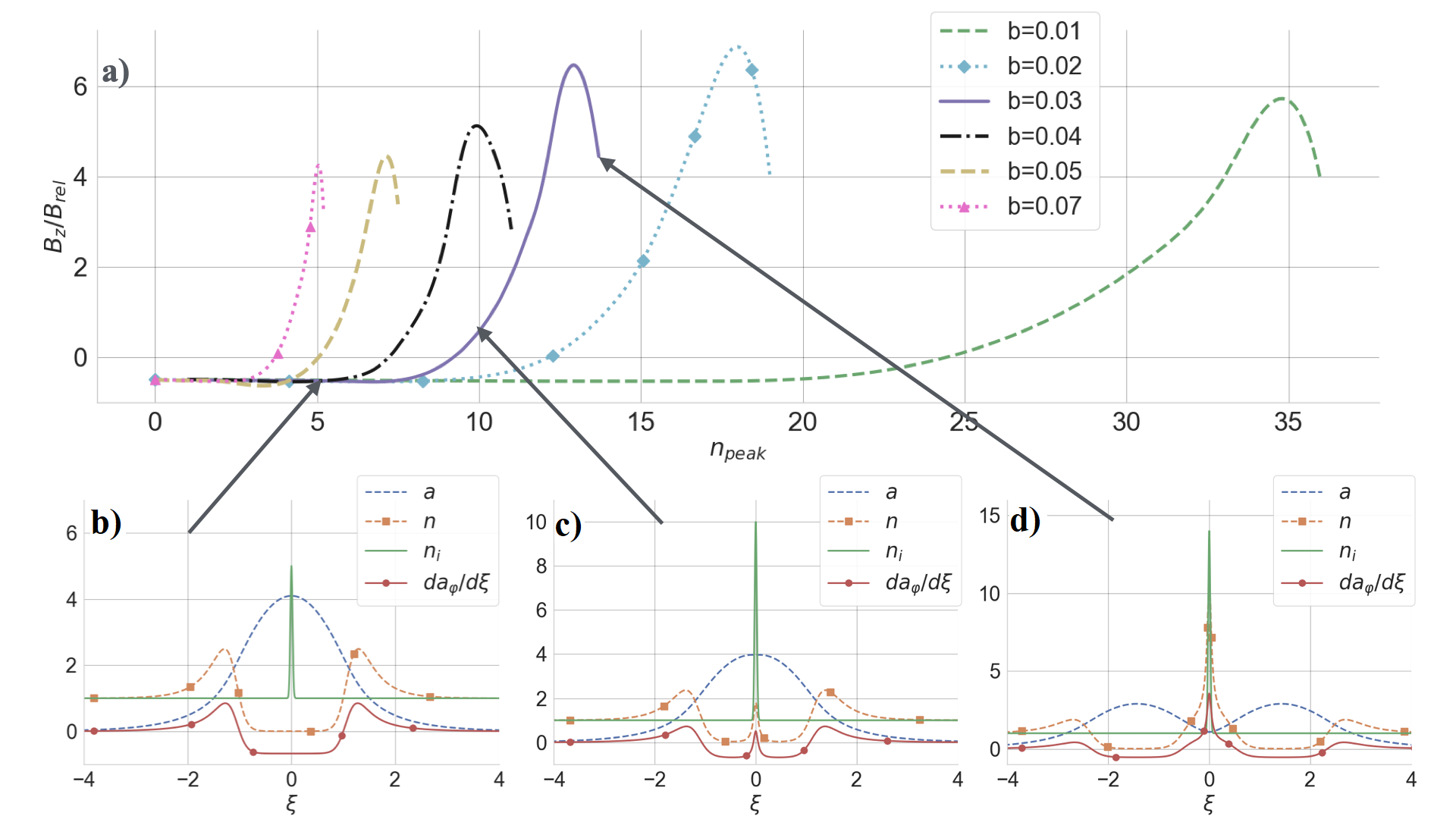}
   \caption{(a) Dependence of the axial magnetic field magnitude on the axis of the laser beam $B_z$ during its self-channeling in a plasma with the transverse density profile (\ref{eq:profile}) on the peak density $n_{peak}$ for different spike widths $b$ at fixed linear power $Q=40 Q_{rel}$ and background concentration $N_0=1.2 N_{cr}^0$. (b)--(d) Examples of distributions of the laser field, electron and ion densities, and axial magnetic field at $n_{peak}=5$, $10$, $14$, respectively, and $b=0.03$. The dependences are plotted for $\mu=0.03$.
   }
   \label{fig:vsPeak}
\end{figure*}

\begin{figure}
   \centering
   \includegraphics[width=\linewidth]{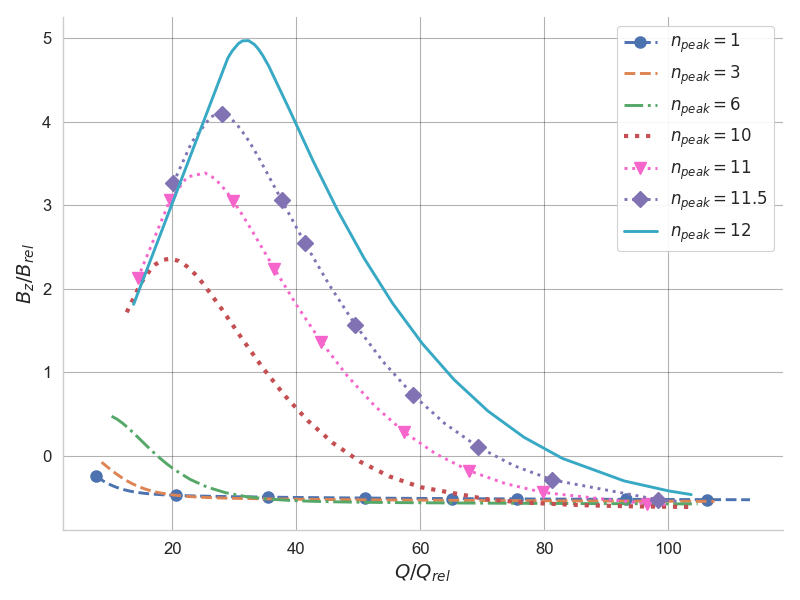}
   \caption{Dependence of the axial magnetic field magnitude on the axis of the laser beam during its self-channeling in a plasma with the transverse density profile (\ref{eq:profile}) on the linear radiation power for the spike width $b=0.03$ and different $n_{peak}$. The dependences are plotted at the background concentration $N_0=1.2N_{cr}^0$ and $\mu=0.03$.
   }
   \label{fig:BvsQ}
\end{figure}

Figure~\ref{fig:vsPeak} shows the dependence of the axial magnetic field on $n_{peak}$ for various spike widths $b$ at fixed background plasma density and radiation power. It can be seen that at small $n_{peak}b$ the magnetic field remains almost unchanged, which is due to the fact that in this regime there are relatively few electrons in the spike and they are expelled by the ponderomotive action of the field (Fig.~\ref{fig:vsPeak} (b)). However, at a certain $n_{peak}b$, the radiation power becomes insufficient to expel all electrons, and a peak in the electron density forms on the beam axis, which leads to the generation of a magnetic field of polarity opposite to the initial one (Fig.~\ref{fig:vsPeak} (c)). This field increases sharply with the number of electrons in the spike, but then begins to decrease because the skin depth becomes smaller than the spike width and the field is expelled from the peak region (Fig.~\ref{fig:vsPeak} (d)). Note that in this regime, the field structure begins to resemble that of the second mode in a homogeneous plasma, with two maxima and a minimum on the beam axis. Nevertheless, this mode is the fundamental one in the sense that it has the lowest power $Q$. The presence of the minimum on the axis is due to diffraction on the plasma density spike.

Thus, there is an optimum in the number of electrons in the spike, at which the maximum magnetic field is achieved. Moreover, the magnitude of this field, as can be seen from Fig.~\ref{fig:vsPeak}, depends weakly on the spike thickness $b$ at sufficiently small values of the latter, which is probably due to the fact that in this case $b$ turns out to be smaller than the Debye radius $\sim \sqrt{\mu/n_{peak}}$, and therefore the width of the peak is mainly determined by the parameter $\mu$. The optimum value is also mainly determined by the product $n_{peak}b$ and, for the considered parameters, lies in the range $n_{peak}b\approx 0.35$--$0.4$.

It is also of interest to investigate how the field in the optimal regime depends on the laser radiation power. To this end, Fig.~\ref{fig:BvsQ} shows the dependence of the generated magnetic field on the linear radiation power for fixed spike parameters. It can be seen that the dependences begin to differ noticeably from the case of a homogeneous plasma only at a sufficiently large number of electrons in the spike. Moreover, for each peak concentration, the maximum of the generated field is achieved at a different power, and this maximum in the optimal regime grows with power almost linearly, in contrast to the case of a homogeneous plasma, in which a saturation of growth was observed above a certain power.

Note also that the magnetic fields obtained in plasma with a density spike exceed the fields in a homogeneous plasma by tens of times and are comparable in magnitude to the amplitude of the laser field: $B/B_{rel} \approx a(0)$. Although this formally violates the applicability of our model, in which the back reaction of the quasi-stationary magnetic field on the electron distribution and the intensity of the laser beam is neglected, taking this effect into account should not strongly affect the estimate of the generated field magnitude. Moreover, the model contains a number of other approximations; therefore, its verification in any case requires comparison with the results of fully kinetic numerical simulations.

\begin{figure}
   \centering
   \includegraphics[width=\linewidth]{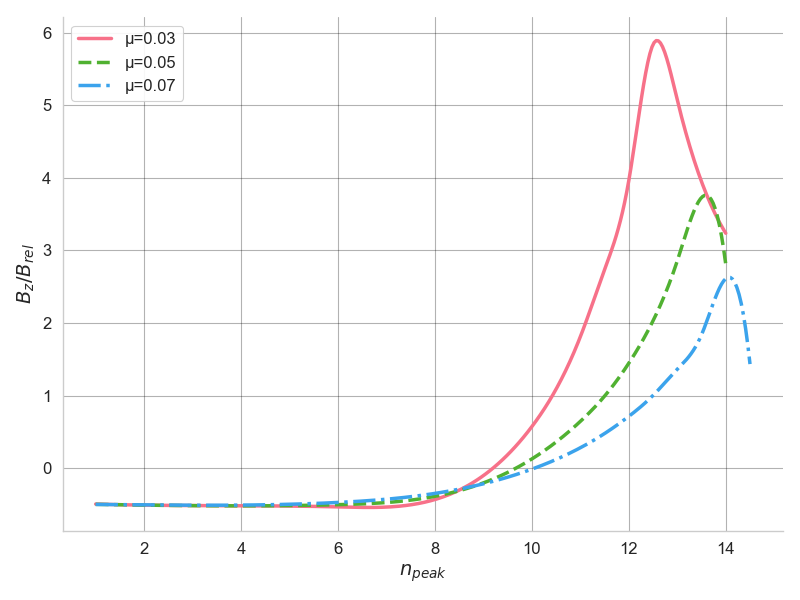}
   \caption{Dependence of the axial magnetic field magnitude on the axis of the laser beam during its self-channeling in a plasma with the transverse density profile (\ref{eq:profile}) on the peak density at fixed linear power $Q=40Q_{rel}$, background concentration $N_0=1.2N_{cr}^0$, and spike width $b=0.03$ for different values of the parameter $\mu$.
   }
   \label{fig:mu}
\end{figure}

Let us also note that, as shown in Fig.~\ref{fig:mu}, the specific values of the generated magnetic field depend significantly on the parameter $\mu$, which, although of a technical nature, has the associated Debye radius scale of the same order as the skin depth. An attempt to decrease the parameter $\mu$ encountered technical difficulties in finding solutions that could not be overcome. Nevertheless, in reality one should expect a term of the same order associated with the electron pressure, since $\mu=0.03$ corresponds to a temperature of the order of $0.03m_ec^2\approx 15$~keV, which is comparable to the characteristic temperatures achievable in laser-plasma experiments due to collisions. Thus, such a simple model does not, generally speaking, allow one to determine the exact values of the generated fields; however, it does allow one to identify the main physical mechanisms of their enhancement through plasma density profiling.

\section{Numerical simulation}

The model considered by us in the previous section allows one to estimate the qualitative behavior of the system as a function of the parameters; however, due to a number of approximations used, it does not allow quantitative estimates to be made. To confirm its main conclusion, namely that the addition of a thin density spike with an optimal number of electrons on the propagation axis of the laser beam makes it possible to generate quasi-stationary magnetic fields with a magnitude comparable to the amplitude of the laser pulse, we have performed numerical simulations.

The simulations were performed using a fully kinetic relativistic electrodynamic code based on the particle-in-cell method, implemented in the PICADOR software package \cite{surmin_CPC_2016}, in three-dimensional geometry in a domain of size $30\times 10\times 10$~$\mu$m$^3$. The plasma was assumed to be preionized and consisted of a cylindrical layer of length 20~$\mu$m in the longitudinal direction and diameter 10~$\mu$m in the transverse direction, centered in the simulation domain. The ions were assumed to be mobile with a charge-to-mass ratio of $e/2m_p$, where $m_p$ is the proton mass, which corresponds, for example, to fully ionized carbon-12 atoms.

The laser pulse with a wavelength of 1~$\mu$m was generated at the left boundary of the simulation domain and had a Gaussian profile in all dimensions. Its duration was 30~fs at the half-maximum intensity level. The pulse was focused into the plane located in the middle of the simulation domain (at a distance of 15~$\mu$m from the left and right boundaries) into a spot of diameter 3~$\mu$m at the half-maximum intensity level. The dimensionless amplitude of the laser pulse at the focus in the absence of plasma was $a_0=10$, which corresponds to an intensity of $1.37\times 10^{20}$~W/cm$^2$.

The background plasma density was fixed and corresponded to an electron concentration of $0.05N_{cr}^0\approx 5.57\times 10^{19}$~cm$^{-3}$. Thus, the critical power for self-focusing was $P_c\approx 350$~GW, and the laser pulse power was $P\approx 120P_c\approx 43$~TW.

The density spike was a uniform cylinder located on the beam axis. The choice of the spike diameter was mainly determined by our computational capabilities for reducing the grid spacing in the transverse dimensions; as a result, it was fixed at 200~nm with a grid spacing of $\Delta y=\Delta z=40$~nm. The grid spacing in the longitudinal dimension was determined by the resolution of the laser pulse wavelength and was chosen to be $\Delta x=100$~nm. The time step was chosen based on the Courant stability condition and was $\Delta t = 0.067$~fs. The number of particles per cell in the background plasma was 1, and in the spike it was proportional to the ratio of the spike density to the background plasma density (6000 for the results presented below).

The plasma density in the spike was varied in order to find the optimal value. With its gradual increase, the longitudinal magnetic field on the system axis increased, but when the electron concentration reached values above $N_\textrm{peak}\approx 300 N_{cr}^0$, the growth practically stopped, and the spike was destroyed during the interaction, which we attribute to the fact that the currents in the spike reached such high magnitudes that the electron Larmor radius became comparable to the spike radius, which led to its destruction.

\begin{figure*}
   \centering
   \includegraphics[width=\linewidth]{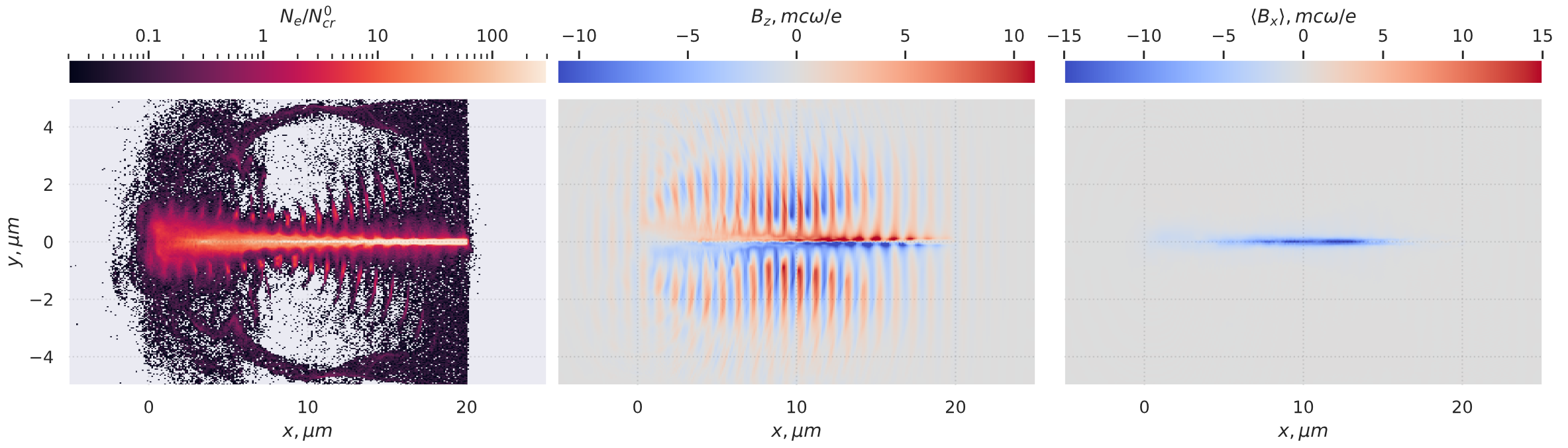}
   \caption{Snapshots of two-dimensional slices in the plane $z=0$ of the electron density (left), the transverse projection of the magnetic field $B_z$ (center), and the longitudinal projection of the magnetic field $B_x$ averaged over the laser period (right), obtained from numerical simulations. The time corresponding to the passage of the maximum of the pulse amplitude through the central plane $x=10$~$\mu$m is shown.
   }
   \label{fig:sim-2d}
\end{figure*}

Figure~\ref{fig:sim-2d} shows the distributions of electrons and two projections of the magnetic field at $N_\textrm{peak}= 300 N_{cr}^0$ at the time corresponding to the passage of the laser pulse maximum through the central plane $x=10$~$\mu$m. It can be seen that in the plasma, under the action of the radiation, a cavitation region is formed, while the pulse undergoes self-focusing. The transverse magnetic field contains both an oscillating component and, near the axis, a quasi-static component associated with the longitudinal acceleration of electrons. At the same time, a strong quasi-stationary longitudinal field is also generated on the axis, the magnitude of which is comparable to and even exceeds the magnitude of the transverse field.

\begin{figure}
   \centering
   \includegraphics[width=\linewidth]{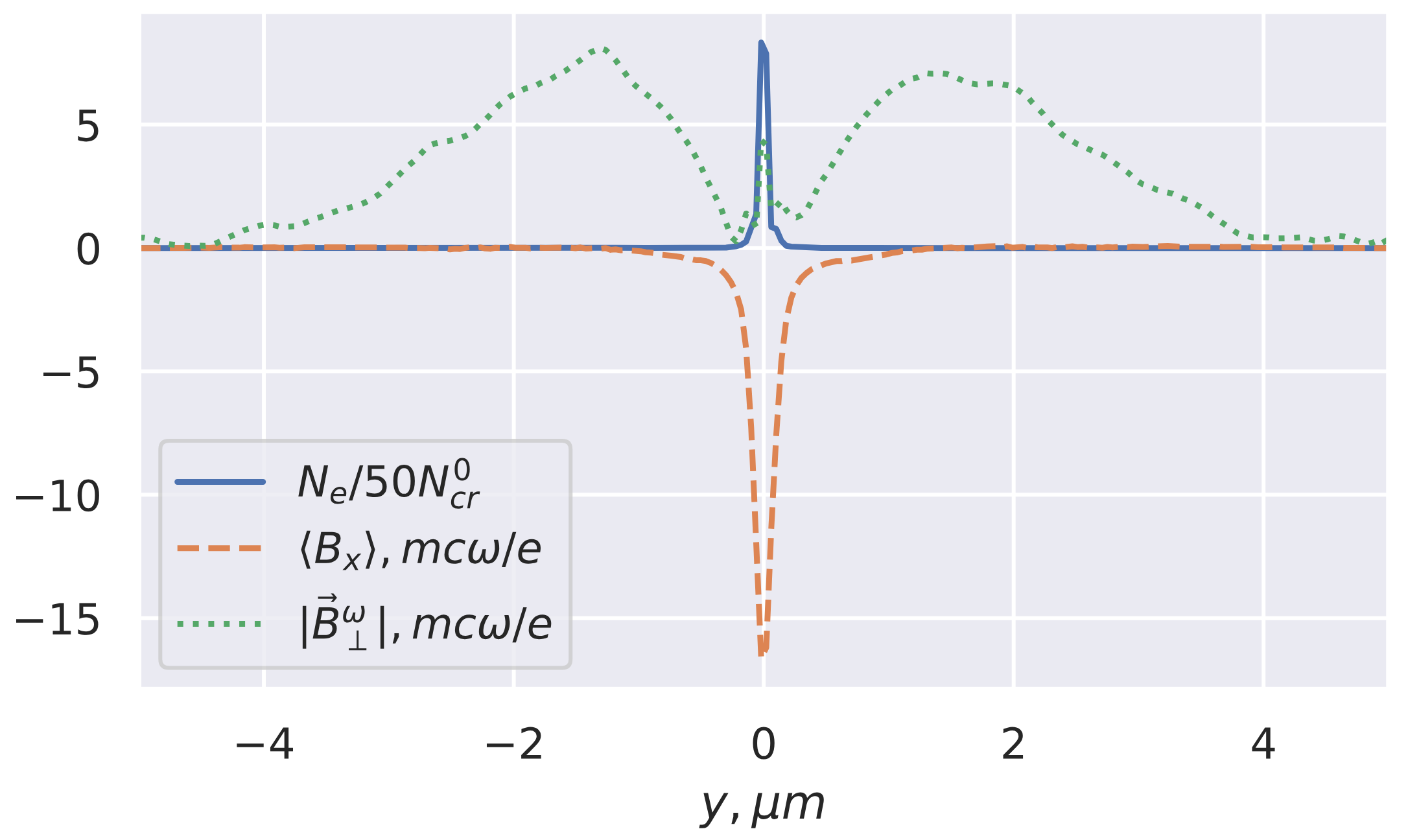}
   \caption{Snapshot of a one-dimensional cut at $z=0$, $x=12$~$\mu$m of the transverse distribution of the electron density, the longitudinal projection of the magnetic field averaged over the laser period, and the oscillating part of the transverse projection of the magnetic field $\left|\vec B_\perp^\omega\right| = \left((B_y-\langle B_y\rangle)^2 + (B_z-\langle B_z\rangle)^2\right)^{1/2}$, where $\langle\cdot\rangle$ denotes averaging over the laser wave period.
   }
   \label{fig:sim-1d}
\end{figure}

Figure~\ref{fig:sim-1d} shows the transverse profiles of the electron density and the transverse and longitudinal fields along the $y$ axis for $z=0$ and $x=12$~$\mu$m, where at this moment the maximum value of the generated quasi-stationary axial magnetic field is reached. As can be seen, its magnitude reaches $16B_{rel}\approx 1.5$~GG, which is more than one and a half times higher than the field amplitude in the laser pulse in the absence of plasma. The transverse field in this case has a characteristic two-hump structure caused by the expulsion of the field from the density spike. Note that inside the spike there is a relatively strong localized mode with a transverse magnetic field, which we associate with a surface electron charge wave excited by the laser pulse and clearly visible in Fig.~\ref{fig:sim-2d} (left). Note also that the electron concentration at the center of the distribution is even higher than initially (about $350N_{cr}^0$), which indicates compression of electrons, presumably due to the ponderomotive action of the laser pulse and the pinch effect of the currents excited in the spike. At the same time, significant ion expansion in the interaction region is not observed, despite the fact that the inverse ion plasma frequency $f_{pi}^{-1}=2m_p/(2\pi e^2\cdot 300N_{cr}^0)\approx 9.1$~fs is noticeably shorter than the duration of the laser pulse.

Thus, realistic numerical simulations, fully taking into account the complex kinetic nature of the interaction of laser radiation with plasma, including ion motion, confirm that the addition of a thin density spike of optimal density on the propagation axis of the laser pulse makes it possible to generate in it a quasi-stationary magnetic field with a magnitude comparable to that of the laser field.

\section{Conclusion}

In this work, we have investigated the possibility of enhancing the generated quasi-stationary magnetic field by ultra-intense laser pulses during their interaction with plasma by adding a relatively thin (of the order of the skin depth in thickness) plasma density spike to the interaction region. Such a spike does not lead to significant diffraction of the radiation, but at the same time increases the number of electrons in the region of the strong field, which makes it possible to excite larger currents in the same volume.

This idea was tested on the model of magnetic field generation via the inverse Faraday effect in the stationary self-channeling regime of a circularly polarized laser beam. It was shown that, in contrast to the case of a homogeneous plasma, in which the growth of the magnetic field with increasing radiation power is strongly suppressed above a certain power, in the case of plasma with a density spike the magnetic field, with an optimal choice of the spike parameters, continues to grow continuously. At the same time, the magnitude of the generated field turns out to be comparable to the amplitude of the laser wave, which in the future makes it possible, for radiation intensities of the order of $10^{26}$~W/cm$^2$ planned in future projects, to expect the generation of fields of the order of a teragauss. At such high intensities, however, it is necessary to take into account additional physical effects, in particular, radiative losses of particles, electron-positron pair production, and ion mobility comparable to that of electrons. This may become the subject of future research.

In previous studies of the generation of both axial and azimuthal magnetic fields, including those using laser pulses carrying orbital angular momentum, the generated fields were typically obtained to be an order of magnitude or more weaker than the laser fields \cite{pukhov_PRL_1996,nakatsutsumi_NC_2018,kim_PRL_2002,naseri_PP_2010a,liseykina_NJP_2016,lecz_SR_2016,nuter_PRE_2018,shi_PRL_2023}. Comparable magnitudes have been demonstrated either through induced target implosion~\cite{murakami_SR_2020} or through pinching of the excited currents~\cite{kaymak_PRL_2016}. These approaches can be combined with our proposed idea of field enhancement and may lead to even stronger fields. This, however, requires further investigation.

In practice, the density spike under discussion can be obtained by preionization with a controlled delay of thin foils (in two-dimensional geometry) or rods (in three-dimensional geometry) followed by their expansion by the time the main pulse arrives. For example, to obtain a cylindrical layer with a density of $n_{peak}=10$, which at $N_0\approx N_{cr}^0$ corresponds to an electron concentration of $N_e\approx10^{22}$~cm$^{-3}$, and a thickness of $b=0.03$, corresponding to approximately 30 nm, it is necessary to ionize a carbon rod with an electron concentration in the solid state of about $4\times 10^{23}$~cm$^{-3}$ and a diameter of 5~nm. Such structures can be fabricated using modern nanostructuring methods.

\begin{acknowledgments}
This research was funded by Russian Science Foundation grant number 22-19-00371.
\end{acknowledgments}

\bibliography{arxiv}

\end{document}